\documentclass{aa}
\usepackage{graphicx}
\usepackage{lscape}

\begin{document}

   \title{Metallicities of the $\beta$ Cephei stars from low-resolution ultraviolet spectra}

   \author{E. Niemczura, J. Daszy\'nska-Daszkiewicz}

   \offprints{E. Niemczura}

   \institute{Astronomical Institute, Wroc{\l}aw University,
             ul. Kopernika 11, 51-622 Wroc{\l}aw, Poland\\
             \email{eniem@astro.uni.wroc.pl}}

   \date{Received ...; accepted ...}

   \abstract{We derive basic stellar parameters (angular diameters, effective temperatures, metallicities)
   and interstellar reddening for all $\beta$ Cephei stars observed during the IUE satellite mission,
   including those belonging to three open clusters. The parameters are derived by means of an algorithmic
   procedure of fitting theoretical flux distributions to the low-resolution IUE spectra and ground-based
   spectrophotometric observations. Since the metallicity has a special importance
   for pulsating B type stars, we focus our attention in particular on this parameter.
   }

   \titlerunning{Metallicities of the $\beta$ Cephei stars from low-resolution UV spectra}
   \authorrunning{Niemczura, Daszy\'nska-Daszkiewicz}
   \maketitle

\section{Introduction}

$\beta$ Cephei variables are a well-known group of early B-type pulsating stars. Their pulsations
are driven by the classical $\kappa$-mechanism, operating in the layer of metal opacity bump
at $T\approx 2\cdot 10^5~{\rm K}$ caused by a huge number of the absorption lines of the
iron-group elements. The explanation of oscillations in these stars was possible only after
the publication of the new opacity tables by Iglesias et al. (\cite{iglesias92}), and was given by
Cox et al. (\cite{cox}), Moskalik \& Dziembowski (\cite{moskalik}), Kiriakidis et al. (\cite{kiriakidis}) 
and Dziembowski \& Pamyatnykh (\cite{dziembowski}). As was shown by Pamyatnykh (\cite{pamyatnykh99}), 
the extent of the theoretical instability domain in the H-R diagram is very sensitive
to the heavy element abundance, $Z$. The instability domain decreases with decreasing value of $Z$,
and for $Z\approx 0.01$ there are no unstable modes in the observed $\beta$ Cephei domain.
Therefore, the determination of metallicities is the first step of testing the theory
of pulsational instability in massive Main Sequence stars.
It would be important to fix the minimum observed value of [m/H] for which the oscillations
are still present in these stars. Most $\beta$ Cep pulsators belong to our Galaxy,
and they are supposed to have the metal abundance close to the solar value.
Recently, Pigulski \& Ko{\l}aczkowski (\cite{pigulskikol02}) discovered the first extragalactic
$\beta$ Cep stars in LMC, where the average metallicity is much below the solar value.
All LMC $\beta$~Cep stars are in or close to young associations and they have probably
higher metal abundance than average.  Moreover, the question "which pulsational parameters
are correlated with the metal abundance ?" is still open.

In this paper we analyze the IUE ({\it International Ultraviolet Explorer}) 
data combined with ground-based spectrophotometric observations of $\beta$ Cephei stars.
The ultraviolet (UV) part of the spectra of main-sequence B-type stars is very rich in lines
of the iron-group elements. Because the greatest amount of energy for these objects is emitted
in the spectral region between 1216 and 10000~\AA, even the low-resolution spectra can
give credible values of stellar parameters such as the effective temperature, $T_{\rm eff}$,
surface gravity, $\log g$, microturbulent velocity, $v_t$, metallicity, [m/H],
angular diameter, $\theta$, and the color excess, $E(B-V)$.
The usefulness of the low-resolution IUE spectra for this purpose has already been
demonstrated by many authors. Fitzpatrick \& Massa (\cite{fitzpatrick99}) obtained the mean stellar
parameters for fourteen main sequence stars in the spectral range O9.5 -- A1.
Morossi et al. (\cite{morossi}) used IUE spectra, combined with the visual spectrophotometric
observations, to derive [m/H], $T_{\rm eff}$ and $\theta$ for 53 objects in the solar
neighborhood. In Niemczura et al.(\cite{niemczura02}) we presented the results for some objects
from the paper of Code et al. (\cite{code}) and for the Magellanic Clouds B type stars, obtained
by the least-squares optimization algorithm.
In a previous paper (Daszy\'nska et al. \cite{daszynska02}), using the above mentioned method,
we determined the values of the metal abundance parameter in the atmospheres of 31 field $\beta$ Cephei stars
and 16 stars belonging to three open clusters NGC~3293, NGC~4755 and NGC~6231.
Here we use the same data as in Daszy\'nska et al. (\cite{daszynska02}) but we derive the stellar parameters
with the improved procedure which allows, in particular, for a more objective error estimation.
This method was applied recently by Niemczura (\cite{niemczura03}) to determine the
metallicities, effective temperatures, stellar diameters and interstellar
reddenings for Slowly Pulsating B-type stars. 

In Sect.~2, all stars we analyze are presented and spectrophotometric observations together with
their calibration are shown. The procedure of fitting the theoretical spectra to the observations
and the method of bootstrap resampling adopted for determining the errors of parameters
are briefly described in Sect.~3. The results and interparametric correlations are presented in Sect.~4.
In Sect.~5, we discuss in detail the values of metallicities we obtained. Conclusions are given in Sect.~6.

\section{The observational material}
The observational material consists of IUE observations obtained with the large or small aperture. 
We connected the data from both the long-wavelength (LWP and LWR, 1950--3200~{\AA}) and the short-wavelength
(SWP, 1100--1950~{\AA}) cameras. We used observations with high (0.1--0.3~{\AA}) and low (6--7~{\AA}) 
spectral resolutions. The data were processed by means of two reduction packages:
IUE/NEWSIPS ({\it New Spectral Image Processing System}, Nichols \& Linsky \cite{nichols},
Gahart et al. \cite{gahart}) and IUE/INES ({\it IUE Newly Extracted Spectra},
Rodr{\'{\i}}guez Pascual et al. \cite{rodriguez}, Gonz\'ales-Riestra et al. \cite{gonzales}
and the references therein). A detailed comparison of the IUE low-dispersion spectra extracted by the INES
and NEWSIPS procedures made by Schartel \& Skillen (\cite{schartel}) showed an excellent
agreement in most cases. The main differences were found for underexposed spectra and strong spectral lines.
Niemczura (\cite{niemczura03}) carried out a comparison of the IUE spectra processed by NEWSIPS and INES packages for 
SPB and reference stars. A similar comparison for the hotter stars made in this work
gives the same results. The systematic differences appear in strong lines and at the wavelengths where
the LWP/LWR and SWP spectra overlap. Large differences also occur in the wavelength range 2000 -- 2300~{\AA}
for highly reddened stars. Other discrepancies are probably not systematic. As found
by Niemczura (\cite{niemczura03}), better agreement exists for high-resolution data.

All data points with a flag indicating some data-quality problems were excluded from further analysis.
Because of the presence of interstellar Ly$\alpha$ absorption, the spectral region around Ly$\alpha$
was omitted.  The IUE observations for $\lambda > 3000$~{\AA} were also excluded from the analysis
because of the very low signal-to-noise ratio at these wavelengths. Thus, we used the IUE fluxes
in the spectral range from 1300 to 3000~{\AA}.  Whenever more than one spectrum was available for a star,
the observations were co-added using a cross-correlation technique. The spectra were re-binned to
theoretical wavelength points.

The ultraviolet observations expressed in absolute units were supplemented by ground-based spectrophotometric
measurements taken from the Breger Catalogue (Breger \cite{breger}), Pulkovo Spectrophotometric Catalogue
(Alekseeva et al. \cite{alekseeva}), and from catalogues of Glushneva et al. (\cite{glushneva98}),
Glushneva et al. (\cite{glushneva92}) and Kharitonov et al. (\cite{kharitonov88}). 
If no spectrophotometric data were available (see Tables~\ref{enjdtab1} and \ref{enjdtab2}),
we used Johnson $UBV$ and Str\"omgren {\it uvby} magnitudes. The magnitudes were converted
into fluxes by means of the formula $F_\lambda = 10^{0.4(C_\lambda - m_\lambda)}$, with scaling factors
$C_\lambda$ taken from Gray (\cite{gray}) and Bessell (\cite{bessel}) for the {\it uvby} and $UBV$ filters,
respectively. All photometric data used in this paper were taken from the GCPD database
({\it General Catalogue of Photometric Data}, Mermilliod et al. \cite{mermilliod}).

All stars selected for the analysis are presented in the Tables~\ref{enjdtab1} and \ref{enjdtab2}, where their names, HD numbers
(ID numbers for cluster stars),
numbers of IUE spectra, parallaxes and references for the ground-based observations of field objects are shown.
Many $\beta$ Cephei stars we analyze belong to binary or multiple systems.
Some of them are wide visual binaries with separation greater than 20~arcsec and can be resolved
by the IUE cameras. In this group we have eight stars:
$\nu$~Eri, 12~Lac, $\beta$~CMa, $\xi^1$~CMa, $\epsilon$~Cen, $\alpha$~Lup, BU~Cir and $\gamma$~Peg.
In some cases the parameters of the companion indicate that it
does not influence the spectrum of the much hotter $\beta$~Cephei star. 

$\tau^1$ Lup is a visual binary with a separation of 158.2~arcsec and a difference in magnitude
of 4.7, therefore the visual component does not influence the spectrum of the $\beta$~Cephei star.
But Proust et al. (\cite{proust}) found a third component with a separation from the primary 
of the order of 0.3~arcsec. We have no information about this object.
27 CMa is a binary system resolved by speckle interferometry
(see for example Hartkopf et al. \cite{hartkopf00}). The separation of the system is growing,
the last determined value being equal to 0.133$\pm$0.008~arcsec. Abt \& Cardona (\cite{abt84}) found the 
orbital period
of the system to be equal to $P_{orb} \approx 40$~y.  $\theta$ Oph  has a close companion with a separation  
of $0.300\pm0.025$~arcsec (Shatsky \& Tokovinin \cite{shatsky}). 
BW Vul is a binary system ($P_{\rm orb} = 33.3\pm$0.3~y, Pigulski \cite{pigulski92a}) with 
the secondary component 6 -- 10~mag fainter than the primary.
EN Lac is an eclipsing binary with orbital period 12.0968~d (Pigulski \& Jerzykiewicz \cite{pigulskijerz88}, 
Lehmann et al. \cite{lehmann01} and references therein). The secondary
component is fainter than the primary by about 5~mag. 
In all above systems there is no effect of the secondary component on the UV spectrum of the $\beta$ Cephei star.
The same is true for $\nu$ Cen, which is an SB system with an orbital period equal to 2.622$\pm$0.018~d
(Schrijvers \& Telting \cite{schrijvers}) and a separation of $5.588\pm0.061$~arcsec (Shatsky \& Tokovinin \cite{shatsky}).
Merezhin (\cite{merezhin94}) determined the masses of the components to be equal to 10.20$\pm$0.10 and 
1.57$\pm$0.25~M$_\odot$. $\sigma$ Sco is a spectroscopic binary (SB2) with a period of 33~d, and a component
of a quadruple system (Pigulski \cite{pigulski92b}, Chapellier et al. \cite{chapellier92}). The separation of the third
component is equal to 0.4~arcsec and is growing (Pigulski \cite{pigulski92b} and references therein).
The brightness of this star is about 2.2~mag lower than that of the SB2 system and its orbital period 
is longer than 100~y (Evans et al. \cite{evans86}). The fourth object in this system is separated
by about 20~arcsec from the primary. $\beta$ Cep is a triple system with a distant visual component
(separation equal to 13.4~arcsec). Gezari et al. (\cite{gezari72}) found a close component, separated from the primary
by about 0.25~arcsec. By 1992 this value decreased to 0.03~arcsec. The orbital elements
of this system were determined by Pigulski \& Boratyn (\cite{pigulskibor92}) who derived a period of 91.6$\pm$3.7~y
from the light-time effect.

In a few cases the spectra in the multiple systems are strongly influenced by the other components.
$\beta$ Cru has a distant visual component (separation of about 371.6~arcsec).
In addition, it is a primary component of a binary system. The orbital parameters of this system
were determined by Aerts et al. (\cite{aerts98}), who derived a period of 1828.0$\pm$2.5~d. The orbital parameters 
allowed to estimate the masses of the components, to be equal to 16 and 10~M$_\odot$. The spectral 
type of the secondary is B2~V, and is consistent with the temperature of 22,000 -- 23,000~K and $\log g$ = 4.0~dex.

There are several spectroscopic binaries with the $\beta$~Cephei variables as the primary components.
$\alpha$ Vir has an orbital period of 4.0145~d and the primary component is the $\beta$ Cephei variable.
The magnitude of the secondary is lower by about 2.0~mag. 
The SB1 system $\beta$ Cen has an orbital period of 1~y (Shobbrook \& Robertson \cite{shobbrook68}), and
in addition a close visual component with a separation  of 1.3~arcsec and $\Delta m = 3.2$~mag.
The separation in the SB1 system is 15.6$\pm$2~mas (Robertson et al. \cite{robertson99}).
Ausseloos et al. (\cite{ausseloos}), on the basis of observations of lines in the visual part of the spectrum,
estimated that both components are $\beta$ Cephei-type stars. $\kappa$ Sco is an SB2 system with orbital 
period of 195~d (Molenberghs et al. \cite{molenberghs}). The orbital parameters of the system and parameters 
of the components were determined by Uytterhoeven et al. (\cite{uytterhoeven}).
Effective temperatures amount to 22,000 and 18,000~K, and the masses are equal to 17 and 12~M$_\odot$, respectively.
$\lambda$ Sco is also a spectroscopic binary. The separation of the system is equal to 5.595~arcsec (De Mey et al. \cite{demey97}).
Heynderickx et al. (\cite{heynderickx94}) estimated effective temperatures (21,200 and 23,100~K), surface gravities
(3.77 and 3.76~dex) and masses (10.81 and 10.50~M$\odot$) of the components. From the analysis
of Bergh{\"o}fer et al. (\cite{berghofer00}) of IUE observations it follows that the second component is a white dwarf.
This SB system also has a visual component with a separation of 1.3~arcsec and $\Delta m$ = 3.2~mag.

\section{The analysis}
The method of analysis was presented in detail by Niemczura (\cite{niemczura03}). Here we give only
brief description of the method. We analyzed low-resolution spectra by means of an algorithmic
procedure of fitting the theoretical flux distributions, $f_\lambda ^t$, to the observations, $f_\lambda$.
The theoretical flux depends on several atmospheric parameters, like $T_{\rm eff}$, [m/H], $\log g$, $v_t$.
Also the angular diameter of the star, $\theta$ and interstellar extinction described by $E(B-V)$
influenced the theoretical spectrum.
We used theoretical fluxes calculated by Kurucz (\cite{kurucz}) with the standard value
of the microturbulent velocity, $v_t = 2$~km~s$^{-1}$. The spectral
resolution of the observational and theoretical data were adjusted before the analysis.
To derive the vector of all parameters, we used the least-squares optimization algorithm
(Bevington \cite{bevington}, Press et al. \cite{press}, Niemczura \cite{niemczura03}).

We adopted the mean interstellar reddening curve of Fitzpatrick (\cite{fitzpatrick}) for
the majority of the analyzed stars. For stars with high reddening ($E(B-V)>0.10$~mag) the mean curve is not
a good approximation because of spatial variability of the extinction law.
In these cases we computed five additional parameters specifying the shape of the UV extinction
curve (Fitzpatrick \cite{fitzpatrick}). Three parameters describe
the Lorentzian-like bump at 2200~{\AA} (its width, $\gamma$, position, $x_0$ and strength, $c_3$),
one is the far-UV curvature term ($c_4$) and the last one is the linear term ($c_2$).

Two additional parameters, $S_1$ and $S_2$, were used to adjust the short- and long-wavelength
parts of the IUE spectra to the visual flux level. We put $S_1 = 1$ outside the IUE/SWP region,
and $S_2 = 1$ outside the wavelength region covered by the long-wavelength IUE observations.
These parameters are included in the analysis for several reasons. First, for some stars we used
the IUE observations made with a small aperture, and these data are not calibrated to the absolute
level with high precision. Second, the flux level changes during the pulsational cycle.
Third, the SWP and LWP (or LWR) observations we used were sometimes made some years apart.
In such cases, errors in absolute levels caused by instrumental effects can be significant.
Finally, different absolute calibrations of the UV and visual parts of the spectrum may
result in a difference of levels between these two spectral ranges.

Thus, the vector of all parameters, ${\vec p}$ determined simultaneously can be expressed as
\begin{displaymath}
\vec p = [T_{\rm eff}, {\rm [m/H]}, \theta, E(B-V), S_1, S_2, [\gamma, x_0, c_2, c_3, c_4]].
\end{displaymath}
We used the least-squares fitting method to determine the vector of parameters, $\vec p$.
For more details see Niemczura (\cite{niemczura03}).

The problem of simultaneous determination of all parameters during the best-fit procedure
can be solved if the parameters produce detectable and different spectral signatures.
Fitzpatrick \& Massa (\cite{fitzpatrick99}) illustrated how atmospheric parameters and $E(B-V)$ affect
the spectrum of a hot star and concluded that the spectral signatures produced by these parameters
are different. One can therefore expect that they could be obtained simultaneously
by a best-fit procedure. The same is not true for $\log g$, which cannot be obtained
in many cases or is determined with a very large error.

In order to determine surface gravities in the same way for all stars, we made 
use of photometric calibrations and stellar evolutionary tracks.
The value of $\log g$ was derived as the mean from the results from five methods, but
gravities lower than 3.00~dex and greater than 4.50~dex were excluded from the mean.
We used three methods that employ Str{\"o}mgren photometry (Napiwotzki et al. \cite{napiwotzki}, Balona \cite{balona},
Dziembowski \& Jerzykiewicz \cite{dziembowski99}) and a method which
uses the Geneva photometry (K\"unzli et al. \cite{kunzli}). In the last method,
we estimated the gravities using the formula obtained from stellar evolutionary tracks.
The models were computed for OPAL opacities with $X = 0.7$ and $Z = 0.02$
without taking into account the effects of rotation and convective overshooting.
To this aim we considered Main Sequence evolutionary tracks for masses
from 7.0 to 16.0~$M_\odot$ with a step of 1~$M_\odot$. We derived the following relation:
\begin{equation}
\log g = -12.5894 + 4.4810 \log T_{\rm{eff}} - 0.7870 \log L/L_\odot,
\end{equation}
with a standard deviation equal to 0.01 dex. We found that the formulae
for $Z = 0.01$ and $Z = 0.02$ give a differences in $\log g$ of the order of 0.01 dex.
The luminosity, $\log L/L_\odot$, depends on effective temperature, stellar diameter and parallax.
Here we used Hipparcos parallaxes (ESA \cite{esa}). During the best-fit procedure the current luminosity
was corrected for the Lutz-Kelker bias (Lutz \& Kelker \cite{lutz}). The correction for the luminosity
can be calculated if $\sigma_\pi/\pi < 0.175$. Consequently, we can determine $\log g$
by means of this method only for stars with $\sigma_\pi/\pi < 0.175$ (see Table~\ref{enjdtab1}).
For the $\beta$~Cephei stars in clusters, the distances $d$ to these clusters were taken
from the literature: for NGC~3293 $d = 2750\pm250$~pc (Baume et al. \cite{baume03}), for NGC~4755 $d = 2100\pm200$~pc
(Sanner et al. \cite{sanner}) and for NGC~6231 $d = 1990\pm200$~pc (Baume et al. \cite{baume99}).
This method of determination of $\log g$ was included into the iteration process, 
and the values of $T_{\rm eff}$ and $\theta$ were calculated during the best-fit procedure.
Systematic errors can be caused
by such effects as the chemical composition $(X,Z)$, convective core overshooting, rotation etc.
For lower masses there may also be a problem with the treatment of stellar convection.

Formal errors of best-fit parameters resulting from the least-squares method
are definitely underestimated.
The technique of bootstrap resampling is probably the most useful method for
estimating of confidence levels for complex least-squares solutions
(Press et al. \cite{press}, Chap. 15.6, see also Maceroni \& Ruci{\'n}ski \cite{maceroni}).
This technique uses the input data set, $D_0$, consisting of $N$ points,
to generate simulated data sets, $D_i$, with the same number of points.
The symbol $i$ denotes the successive simulation.
The number of all bootstrap simulations is equal to $N(\log N)^2$ (Babu \& Singh \cite{babu}).
In the generated data, the random fraction of original points (37\% in our case)
is replaced by the remaining  original points by means of random resampling with replacement.
The simulated data sets are analyzed in the same way as the original data.
This procedure not only makes it possible to estimate reliable uncertainties of the
parameters, but it can be also used for determining correlations between them.

\section{Results}
\subsection{The field $\beta$~Cephei stars}
   \begin{figure*}
   \centering
   \includegraphics{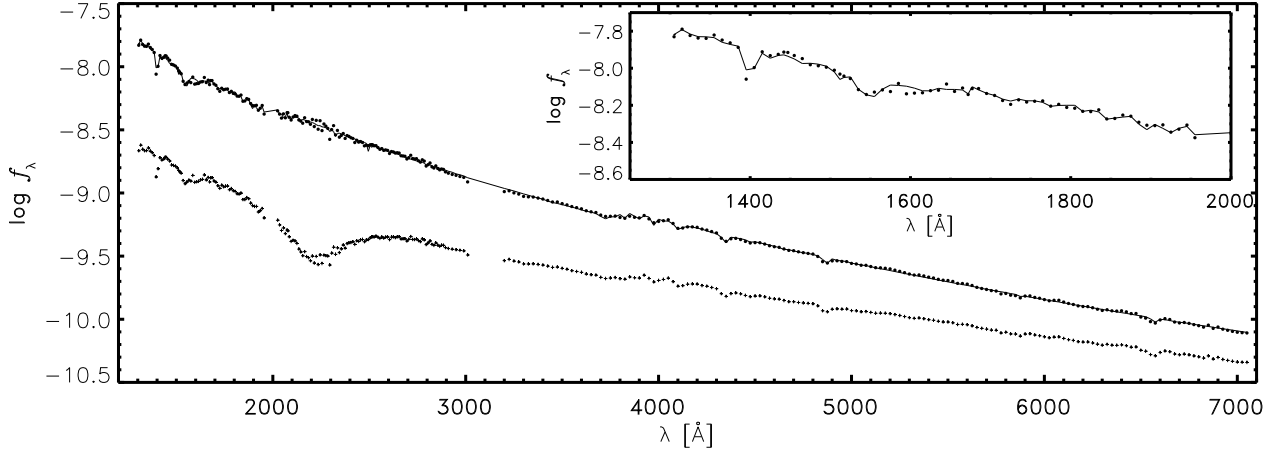}
   \caption{Dereddened stellar energy distribution of $\omega^1$~Sco ({\it dotted line}) in comparison with 
            the best-fit model ({\it solid line}). The original flux of $\omega^1$~Sco is also shown ({\it crosses}). 
            The IUE data were calibrated by means of the NEWSIPS reduction package. 
            In the inset, there is a closer view of the short-wavelength part of the spectrum. }
   \label{enjd1}
   \end{figure*}

   \begin{figure*}
   \centering
   \includegraphics{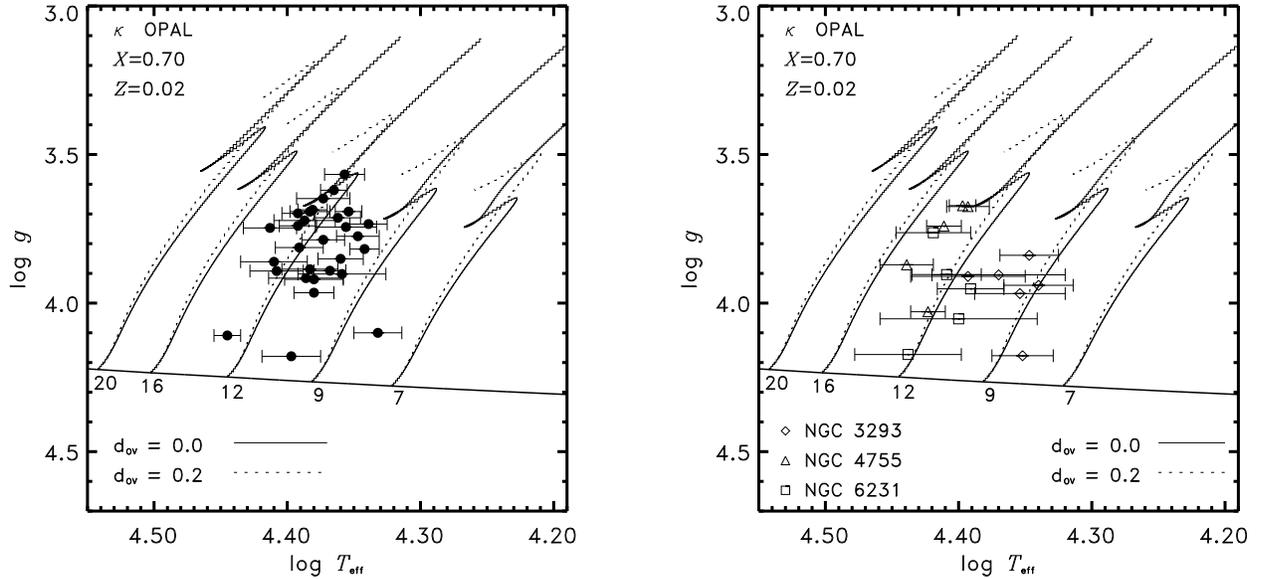}
   \caption{ The position of the field $\beta$ Cep stars (left panel) and those from the three open
             clusters (right panel) on the $\log g$ vs. $\log T_{\rm eff}$ diagram. The evolutionary
             tracks were calculated without (solid line) and with (dotted line) 
             taking into account the effect of convective core overshooting. We assumed $d_{\rm ov} = 0.2$.}
   \label{enjd2}
   \end{figure*}

The parameters obtained for the field $\beta$~Cephei stars from the IUE/NEWSIPS spectra 
analyzed by the bootstrap resampling method are presented in Table~\ref{enjdtab3}. For a few stars 
($\delta$~Cet, 19~Mon and $\delta$~Lup) the differences between the parameters obtained on the basis of the IUE 
data calibrated by the NEWSIPS and INES procedures exceed the error bars.
In the case of $\delta$~Lup the main reason for the large discrepancy can be the low quality of the IUE data
because there is only a single SWP spectrum and a single LWP/LWR spectrum. There is also only one LWP/LWR spectrum
for 19~Mon. In Fig.~\ref{enjd1} we show a comparison of the dereddened
spectrum of $\omega^1$~Sco and the theoretical spectrum resulting from
the bootstrap resampling method. The theoretical spectrum was obtained for
the parameters listed in Table~\ref{enjdtab3}. The data are plotted as a dotted line, the theoretical spectrum
as a solid line. As one can see, there is a good overall agreement between these two spectra.
In the inset, a closer view of the SWP part of the IUE/NEWSIPS spectrum is shown.
All features of the observed spectrum are present in the theoretical fit. 

Most stars we analyzed have small interstellar reddening. For stars with $E(B-V) > 0.1$~mag,
the five additional parameters describing the extinction curve were determined (see Sect.~3)
and are given in Table~\ref{enjdtab4}. For a comparison, parameters of the standard extinction curve
of Fitzpatrick (\cite{fitzpatrick}) are also shown.

In Fig.~\ref{enjd2} we show the $\beta$~Cephei stars we analyzed in the $\log g$ -- $\log T_{\rm eff}$ diagram.
Evolutionary tracks for $M =$ 7, 9, 12, 16 and 20~$M_\odot$ are also plotted. The stellar models were 
computed using a standard evolutionary code written by Paczy\'nski (\cite{pacz69}, \cite{pacz70})
and improved by Sienkiewicz, Koz{\l}owski and Pamyatnykh (private communication).
We used the most recent version of OPAL opacity data (Iglesias \& Rogers~\cite{iglesias96}).
The effect of rotation was not taken into account. An initial hydrogen abundance
$X = 0.70$ and a metal abundance $Z = 0.02$ were assumed. Two types of evolutionary tracks are shown:
without the effect of convective core overshooting (solid lines), and with the overshooting effect,
assuming $d_{\rm ov} = 0.20$~$H_P$ (dotted lines). As one can see, all $\beta$~Cephei stars
are located inside the Main Sequence area.

\subsection{$\beta$~Cephei stars in open clusters}
The parameters obtained for $\beta$~Cephei stars belonging to the galactic open clusters NGC~3293, NGC~4755 
and NGC~6231 are shown in Table~\ref{enjdtab3}. The analysis was performed for six members of
NGC~3293, five members of NGC~4755, and five of NGC~6231. 
For a few stars (NGC~3293-14, NGC~4755-7, NGC~4755-10, NGC~4755-228, NGC~4755-332, NGC~6231-238, NGC~6231-937)
the values of parameters obtained from the two sets of data, IUE/NEWSIPS and IUE/INES, do not agree
to within their errors. This is mainly caused by the poor quality of the IUE data. 

The reddening is in all cases greater than $0.10$~mag, so that there is a need to use a specified extinction curve. 
The extinction curves for stars belonging to NGC~3293 and NGC~6231 were taken from Massa \& Fitzpatrick (\cite{massa88}), 
while for NGC~4755 the standard extinction curve of Fitzpatrick (\cite{fitzpatrick}) was used.

The study of $\beta$~Cephei stars in open clusters is important because their location in the 
colour-magnitude diagram indicates their evolutionary state. There are eleven open clusters in which
$\beta$~Cephei variables were discovered (a recent example: three $\beta$~Cep stars in 
NGC~6910, Ko{\l}aczkowski et al. \cite{kolaczkowski04}). A still unexplained problem is why some young clusters contain
those variables and others, like NGC~2362 (Balona \& Laney \cite{balonalaney}), do not. 
The other problem is the difference in location of the instability  strips for various clusters
of nearly the same age. According to Balona \& Koen (\cite{balonakoen}), the location of the $\beta$~Cephei 
variables belonging to NGC~4755 on the HR diagram is shifted to lower effective temperatures 
compared to NGC~3293. However, we obtained opposite results: the location of $\beta$~Cephei stars in NGC~4755
is shifted to the higher effective temperatures as compared to NGC~3293. In the case of NGC~4755 we found 
lower values of [m/H] than in the case of NGC~3293. For the lower metallicity the ZAMS line is shifted to the left
in the H-R diagram, thus our results are fully consistent.

\subsection{Correlations between the parameters}
The mean correlations, $\rho_{\rm mean}$, their standard deviations, $\sigma_\rho$, median values,
$\rho_m$, and ranges of correlation coefficients for the $\beta$~Cephei stars are presented in Table~\ref{enjdtab5}.
Because of the subject of this work, the most interesting are the correlations between metallicity and other parameters
obtained simultaneously. The values of $\rho_{\rm mean}$ and $\rho_m$ for metallicity are small and amount to
about 0.2. Also the standard deviations are small, which means that even for individual objects 
correlations are not large. Small values of $\rho_{\rm mean}$ indicate that the metal abundance, [m/H], is not
correlated with the other parameters and can be determined from the best-fit procedure with good precision.
The correlations between the other parameters are larger. The effective temperature is strongly
correlated with the interstellar reddening, $\rho_{\rm mean} = 0.73$. A lower value of $\rho_{\rm mean}$
was obtained for effective temperature and stellar diameter. The smallest correlation was found between
$\log T_{\rm eff}$ and [m/H]. The correlations can be strong for individual objects.
Somewhat lower values of $\rho_{\rm mean}$ and $\rho_m$ are found between $\theta$ and
other parameters; the largest value of the mean correlation is obtained for
$\theta$ and $\log T_{\rm eff}$ ($0.66$), and the lowest one, as in the other cases, between $\theta$ and [m/H]
($0.23$). The correlations between interstellar reddening and others parameters are lower in the last case.
The largest correlation was obtained between $E(B-V)$ and $\log T_{\rm eff}$, while the lowest between 
$E(B-V)$ and [m/H].

   \begin{figure}
   \centering
   \includegraphics{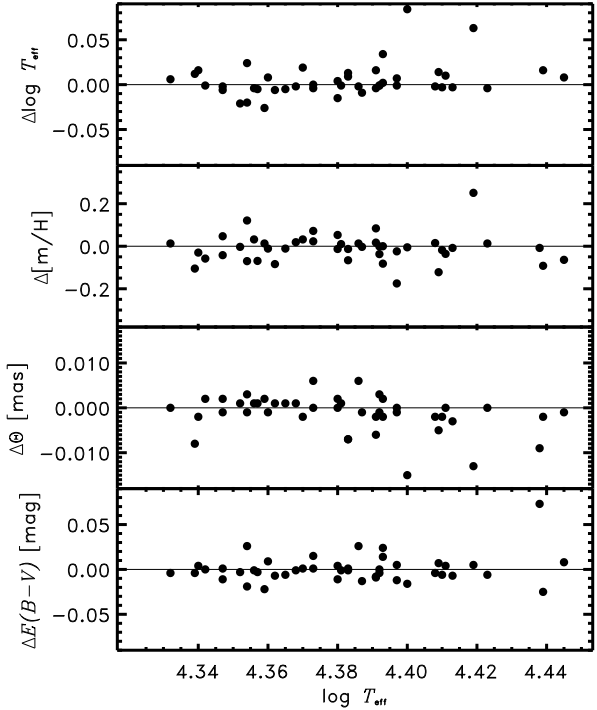}
   \caption{Differences in parameters, $\Delta p_i = p_i({\rm NEWSIPS}) - p_i({\rm INES})$,
            as a function of the effective temperature obtained from the INE/NEWSIPS data.}
   \label{enjd3}
   \end{figure}

   \begin{figure}
   \centering
   \includegraphics{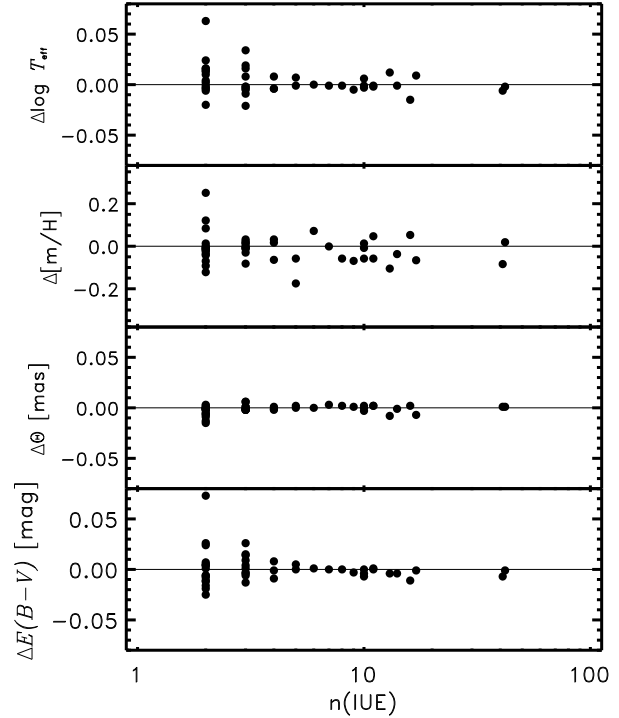}
   \caption{The same as in Fig. 3, but as a function of the number of IUE spectra, n(IUE). }
   \label{enjd4}
   \end{figure}

\subsection{Comparison between NEWSIPS and INES}
In Fig.~\ref{enjd3} we plot the differences between the parameters obtained from the IUE data
calibrated with the NEWSIPS and INES procedures, $\Delta p_i = p_i({\rm NEWSIPS}) - p_i({\rm INES})$,
as a function of $\log T_{\rm eff}$.
In the four panels we can see these differences for $\log T_{\rm eff}$, $\theta$, [m/H] and $E(B-V)$.
For the small range of effective temperature shown in the Fig.~\ref{enjd3} there is no correlation between 
$\Delta p_i$ and $\log T_{\rm eff}$.

Small mean differences, $<|\Delta {\bf p}|>$, were found for all parameters, that is
$0.014\pm0.001$ for $\log T_{\rm eff}$ ($\sigma = 0.023$),
$0.003\pm0.0001$ ($\sigma = 0.003$) for $\theta$, $0.046\pm0.001$ ($\sigma = 0.050$) for [m/H] and 
$0.010\pm0.001$ ($\sigma = 0.012$) for $E(B-V)$. For all this mean values, 
the standard deviations are quite large. This means that the differences for individual objects can be significant.

In Fig.~\ref{enjd4} we show the same differences, $\Delta p_i$, as a function of the number of
available IUE spectra. As one could expect, the largest differences are obtained for stars
with at most two spectra available from each IUE camera. When the number of the IUE spectra grows, the scatter
of $\Delta {\bf p}$ decreases.

\section{Discussion of the parameter {\rm [m/H]}}
\subsection{General properties}
   \begin{figure}
   \centering
   \includegraphics{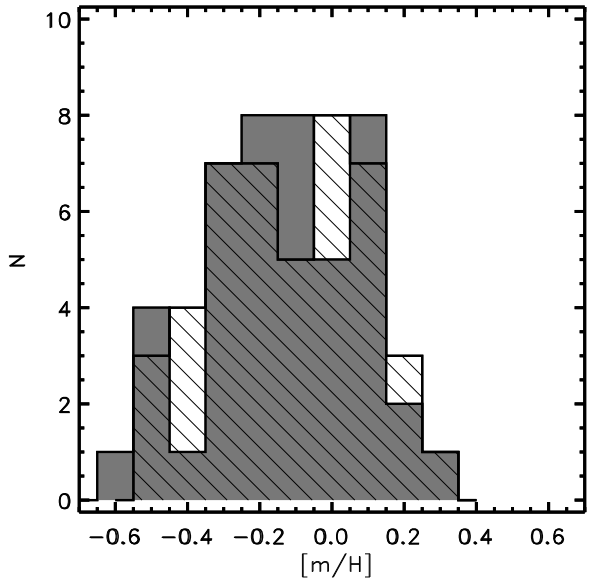}
   \caption{The distribution of the $\beta$~Cephei stars with the [m/H] parameter. The hatched histogram shows the 
            results obtained from the IUE/NEWSIPS data, while the gray one presents results obtained 
            from the IUE/INES data.}
   \label{enjd5}
   \end{figure}

The metallicities of the $\beta$~Cephei stars determined from the IUE/NEWSIPS spectra range from $-0.47\pm0.18$~dex
for 27~CMa to $+0.21\pm0.09$~dex for HN~Aqr. The extreme values determined from the IUE/INES data were obtained
for $\tau^1$~Lup and, again, HN~Aqr; they are equal to $-0.42\pm0.17$~dex and $+0.20\pm0.10$~dex, respectively.
The distribution of the [m/H] parameter resulting from NEWSIPS and INES spectra is shown in Fig.~\ref{enjd5}.
These two samples follow normal distributions within the error limits. Applying the one-way Anova statistical test we obtained
that the two means are not significantly different up to the significance level of $\alpha = 0.83$.
Thus we conclude that there is no 
difference between the values of [m/H] found from the IUE data processed by both reduction packages.
The mean value of the metal abundance parameter for the $\beta$~Cephei stars is equal to $-0.13\pm0.03$~dex
and $-0.14\pm0.03$~dex for IUE/NEWSIPS and IUE/INES, respectively.
The mean metallicity for the field $\beta$~Cephei stars is equal to $-0.14\pm0.03$~dex (IUE/NEWSIPS)
and  $-0.13\pm0.03$~dex (IUE/INES). The mean values of the metal abundance parameter for the cluster
stars, obtained from the analysis of IUE/NEWSIPS spectra is equal to $+0.05\pm0.06$~dex for NGC~3293,
$-0.43\pm0.05$~dex for NGC~4755 and $-0.01\pm0.06$~dex for NGC~6231. 
From the IUE/INES spectra we obtained nearly the same values: $+0.08\pm0.04$~dex for NGC~3293,
$-0.40\pm0.06$~dex for NGC~4755 and $-0.05\pm0.05$~dex for NGC~6231.
These determinations are consistent with our previous results (Daszy\'nska et al. \cite{daszynska02}),
obtained with a less accurate method.

   \begin{figure}
   \centering
   \includegraphics{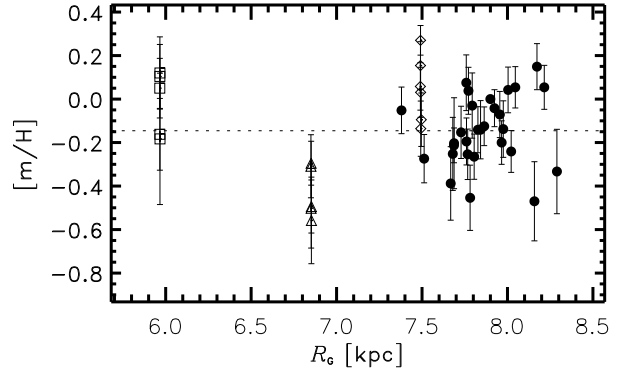}
   \caption{The distribution of the metallicity parameter of the $\beta$~Cephei stars from the IUE/NEWSIPS data as 
            a function of the galactocentric distance (in kpc). The dashed line indicates the mean value of  
            [m/H] obtained for all $\beta$~Cephei stars. Symbols are the same as in Fig.~\ref{enjd2}.}
   \label{enjd6}
   \end{figure}

Metal abundances of hot stars in the solar vicinity lower by about $0.20$~dex than the solar value were reported by
Gies \& Lambert (\cite{gies}, hereafter GL92), Cunha \& Lambert (\cite{cunha}), Kilian (\cite{kilian}), 
Kilian et al. (\cite{kilian94}),
Daflon et al. (\cite{daflon}), Gummersbach et al. (\cite{gummersbach}), Niemczura~(\cite{niemczura03}), and others.
These results were obtained from both high- and low-resolution data. All quoted authors used Kurucz's ATLAS models of stellar 
atmospheres.  Metal abundances obtained in the present paper from the low-resolution IUE
observations are in agreement with these results.

In Fig.~\ref{enjd6} we present the dependence of [m/H] on the galactocentric distance.
The distance for the field stars were calculated by using the Hipparcos parallaxes (ESA \cite{esa}). 
For the stars from open clusters we assumed the same values of the distance from the Sun as in Chapter 3.
The galactocentric distance of the Sun, $R_{G,\odot} = 7.9$~kpc was adopted from McNamara et al. (\cite{mcnamara}).
As can be seen from the figure,
the field stars are spread over a small range of about $0.5$~kpc in galactocentric distance,
so that no clear indication of a metallicity gradient is apparent. However, we would like to point out that
some intrinsic scatter of the metallicity is certainly present. 

\subsection{Comparison with the literature}

For eleven $\beta$~Cephei stars in our sample, the metallicity has been determined by other authors.

One of the well-studied $\beta$~Cephei stars is $\gamma$~Peg. The majority of determinations was obtained from 
the visual part of the spectrum. Snijders (\cite{snijders}) determined [Fe/H] to be equal to $+0.48$~dex, 
while Peters (\cite{peters76}) found a much lower iron abundance, equal to $+0.04$~dex. GL92 obtained [FeII/H]$ = +0.29$~dex
and Pintado \& Adelman (\cite{pintadoadelman}) found [Fe/H]$ = +0.04$~dex. From the high-resolution ultraviolet IUE spectra,
Proffitt \& Quigley (\cite{proffit01}, hereafter PQ2001) determined [Fe/H]$ = -0.15$~dex. The values of [m/H] obtained by us,
$-0.04\pm0.09$~dex and $+0.02\pm0.10$~dex,  
are then consistent with the determinations of Peters (\cite{peters76}) and Pintado \& Adelman (\cite{pintadoadelman}).

The differences between the values of metal abundances for For 16~Lac determined by different authors are significant.
GL92 obtained the iron abundance equal to $-0.23$~dex from the visual part of the spectrum, while from the UV spectrum
PQ2001 obtained [Fe/H]$ = -0.09$~dex, and Venn et al. (\cite{venn}) found [m/H] = $+0.16\pm0.18$~dex. 
Recently, Thoul et al. (\cite{thoul}) determined the abundances of the elements from the visual part of the spectrum. 
According to these authors, the iron abundance amounts to $-0.08\pm0.09$~dex. Our values are 
consistent with the last results to within the errors of determination: 
[m/H] = $-0.14\pm0.13$~dex and $-0.15\pm0.14$~dex.

For $\delta$~Cet GL92 determined the iron abundance to be equal to $-0.24$~dex from the analysis
of the visual part of the spectrum. Different values were obtained from the ultraviolet spectra. From the 
IUE spectra PQ2001 determined [Fe/H] to be equal to $-0.15$~dex.
Venn et al. (\cite{venn}), from HST/STIS spectrograph data, found [m/H]$ = -0.15\pm0.15$~dex. 
All this determinations are in agreement with the present work.
For 15~CMa and  $\beta$~Cep the metal abundances were determined by the same authors. 
GL92 found [Fe/H]$ = -0.47$~dex from the visual part, while from the analysis of the UV data,
PQ2001 and Venn et al. (\cite{venn}) obtained $-0.13$~dex and $+0.10\pm0.10$~dex, respectively.
The values we found are close to the latter determinations.
The metal abundances derived for $\beta$~Cep range from $-0.31$~dex
(PQ2001) to $-0.16\pm0.23$~dex (Venn et al. \cite{venn}). Both these values 
were determined from the UV spectra. From the lines in the visual part of the spectrum, GL92
obtained [Fe/H]$ = -0.23$~dex. The values of [m/H] obtained by us are equal to $-0.07\pm0.11$~dex
and $-0.01\pm0.09$~dex, respectively.

For the rest of the $\beta$~Cephei stars there are only two or one determination of the metal abundances in the literature.
Recently, Stankov et al. (\cite{stankov}) announced a determination of the abundances of elements for BW~Vul, derived from 
high-resolution spectra in the blue.  According to these authors, the iron abundance, [Fe/H], 
amounts to $+0.02$~dex. This is close to the value obtained by us, 
[m/H]$ = +0.09\pm0.02$~dex.

For PHL~346, the first abundance determination was done by Ryans et al. (\cite{ryans}). From visual part of the spectrum,
the authors obtained [Fe/H] = $-0.84$~dex. Ramspeck et al. (\cite{ramspeck}), from the same spectral range, determined a much higher
value of the iron abundance, [Fe/H]$ = -0.12$~dex. In this work, [m/H]$ = +0.21\pm0.09$~dex (IUE/NEWSIPS) and $+0.20\pm0.10$~dex
(IUE/INES) were obtained. The metallicities obtained in the present paper for $\beta$~Cru are consistent with
the values given by Kilian (\cite{kilian94a}), [Fe/H] = $-0.08$~dex (from the visual part of the spectrum) and PQ2001,
[Fe/H]$ = -0.25$~dex. For BU~Cir, the values determined in this work 
are lower than the iron abundance, [Fe/H]$ = -0.12$~dex, given by Kilian (\cite{kilian94a}).
For the following four stars, the values obtained from the visual part of the spectrum are taken from GL92,
while the UV data were analyzed by PQ2001. Our metallicities of $\nu$~Eri 
are close to the [Fe/H]$ = -0.08$~dex given by
GL92, but the iron abundance determined by PQ2001 is lower, namely $-0.27$~dex. 

For $\xi ^1$~CMa we obtained metallicities 
close to the iron abundance of $-0.30$~dex determined by PQ2001.
The values obtained in this paper are also consistent within the error limits with [Fe/H] = $-0.18$~dex obtained by GL92. 
For $\beta$~CMa we obtained [m/H]$ = +0.04\pm0.11$~dex and $+0.08\pm0.11$~dex from the IUE/NEWSIPS and IUE/INES data,
respectively. However, GL92 determined [Fe/H]$ = -0.41$~dex and PQ2001 found that [Fe/H]$ = -0.22$~dex. 
In the case of 12~Lac, the values determined in this paper: $-0.20\pm0.10$~dex (IUE/NEWSIPS) and $-0.22\pm0.10$~dex
(IUE/INES) are higher than the iron abundance obtained by GL92, $-0.41$~dex. The iron abundance found by
PQ2001 is equal to $-0.08$~dex.

\subsection{Metallicity and pulsations}
\begin{figure}
   \centering
   \includegraphics{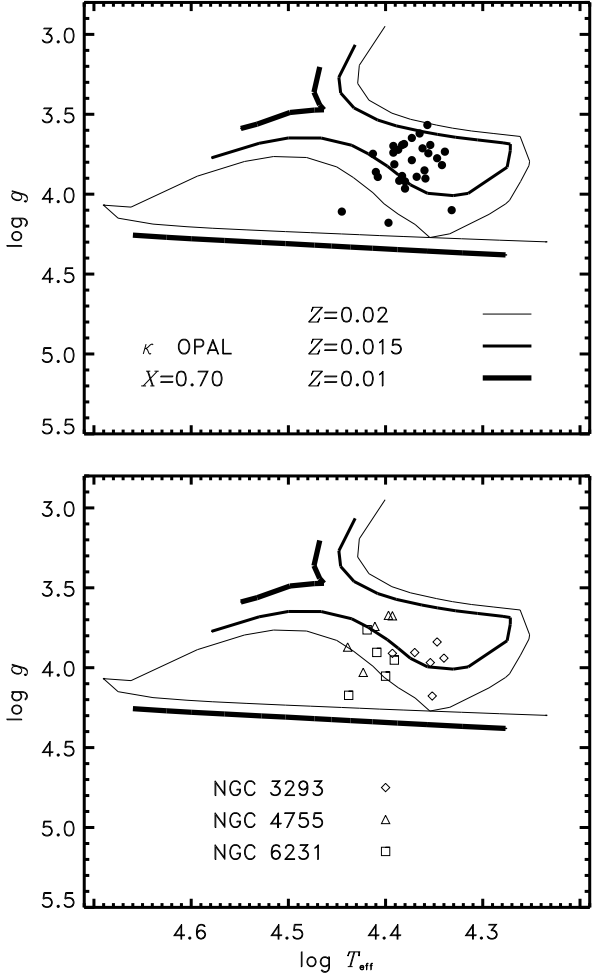}
   \caption{The location of the $\beta$~Cephei stars on the $\log g$ vs. $\log T_{\rm eff}$ diagrams.
            The instability regions for $Z = 0.02$, $0.015$ and $0.01$ were taken from Pamyatnykh (1999). 
            In the upper panel the position of the field $\beta$~Cephei stars is shown, while in the lower
            one, the position of variables from the open clusters.}
   \label{enjd7}
   \end{figure}

In Fig.~\ref{enjd7} we show the instability domain of $\beta$~Cephei variables for three values
of the metal abundance, $Z = 0.01,~0.015,~0.02$, and the stars we analyzed. 
As before, the field $\beta$~Cep stars and those belonging to the three open clusters are plotted in separated panels.
The majority of field $\beta$~Cephei stars is located within the instability region defined by $Z = 0.015$.
Only for three stars are the values of surface gravity higher than 4.0~dex. Two stars, 27~CMa and
$\omega^1$~Sco, are located outside the instability regions. The metallicity parameters, [m/H],
for both are lower than $-0.40$~dex. This is not a surprise as 27 CMa is a Be star. In the case of $\omega^1$ Sco
there is also the possibility that it is a $\zeta$~Oph-type variable. 

\begin{figure}
   \centering
   \includegraphics{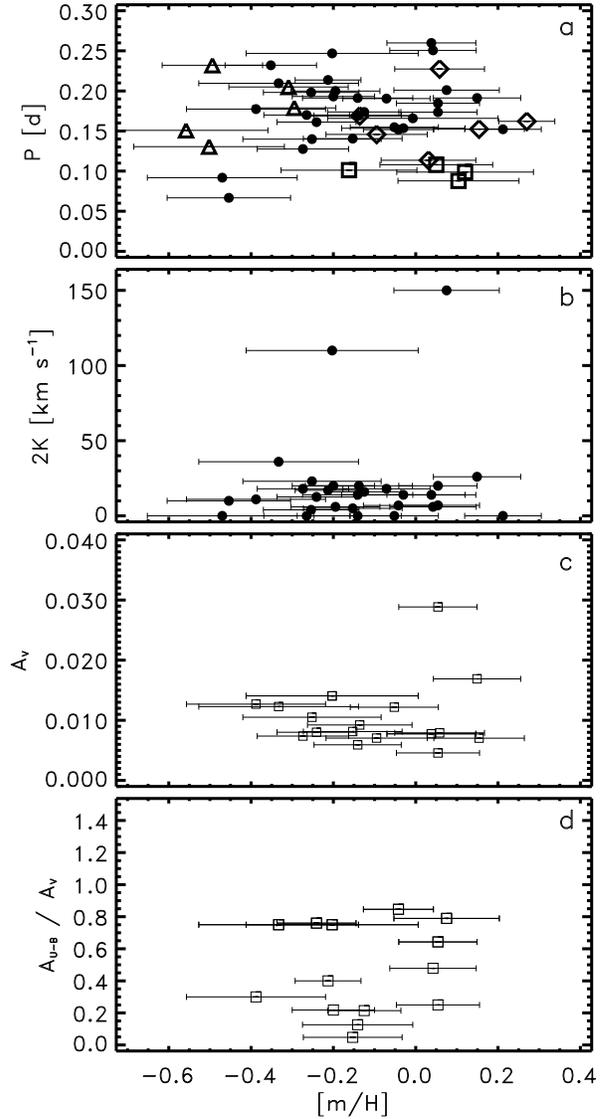}
   \caption{Dependence of pulsational parameters: $P$, $2K$, $A_{\rm V}$ and $A_{\rm U-B}/A_{\rm V}$
            on the metallicity parameter, [m/H].}
   \label{enjd8}
   \end{figure}

To check which pulsational parameter is sensitive to the metallicity in $\beta$ Cep stars,
in Fig.~\ref{enjd8} we plot the period of the dominant mode, $P$, radial velocity amplitude, $2K$, 
its light amplitude in the $V$ filter, $A_{\rm V}$, and a ratio of photometric amplitudes, $A_{\rm U-B}/A_{\rm V}$,
as a function of the [m/H] parameter. We note that the last quantity is independent of the inclination angle, $i$,
and the intrinsic pulsational amplitude, $\varepsilon$. We have checked all passbands and colors available
in the literature, and here we give representative examples.
Panel {\it a} of Fig.~\ref{enjd8} is for the pulsational periods, the values of which are marked as filled circles
for the field $\beta$~Cep stars, and as diamonds, triangles and squares for $\beta$~Cephei stars in the clusters
NGC~3292, NGC~4755 and NGC~6231, respectively. 
As one can see, there is no correlation between metallicity and mentioned pulsational parameters,
as was already pointed out by Daszy\'nska et al. (\cite{daszynska02}). 

Also the projected rotational velocity, $v_{\rm e}\sin i$, is not correlated with the metallicity parameter.

In our previous papers (Daszy{\'n}ska 2001, Daszy\'nska et al. 2003) we stated that the mean metallicity parameter, [m/H],
is higher in the case of multiperiodic $\beta$~Cep stars than in the case of monoperiodic variables. Here we confirm this result.
The mean value of the [m/H] parameter for multiperiodic $\beta$~Cep stars amounts to $-0.07\pm0.03$~dex, 
while the mean metallicity for monoperiodic variables is equal to $-0.26\pm0.06$~dex.
This result is in agreement with the nonadiabatic pulsational theory which predicts a larger number of unstable modes for
higher values of the metal abundance.

The dichotomy of $\beta$~Cep stars with regard to metallicity is discussed in detail in a separate paper 
(Daszy\'nska-Daszkiewicz \& Niemczura \cite{daszynska04}).

\section{Conclusions and prospects}

We derived basic stellar parameters such as the effective temperature, gravity, angular diameter, metallicity
and interstellar reddening for all $\beta$ Cep stars monitored by the International Ultraviolet Explorer
satellite. Among them there were field $\beta$ Cep stars and variables belonging to three
young open clusters: NGC 3293, NGC 4755 and NGC 6231. 
The mean values of the [m/H] parameter for the $\beta$~Cephei stars are equal to $-0.13\pm0.03$~dex
and $-0.14\pm0.03$~dex for IUE/NEWSIPS and IUE/INES data, respectively.
The mean metallicity for the field $\beta$~Cephei variables is equal to $-0.14\pm0.03$~dex (IUE/NEWSIPS)
and  $-0.13\pm0.03$~dex (IUE/INES), while the mean values of [m/H] for the cluster
stars obtained from the analysis of IUE/NEWSIPS spectra are equal to $+0.05\pm0.06$~dex for NGC~3293,
$-0.43\pm0.05$~dex for NGC~4755 and $-0.01\pm0.06$~dex for NGC~6231.
From the IUE/INES spectra, we obtained nearly the same values: $+0.08\pm0.04$~dex for NGC~3293,
$-0.40\pm0.06$~dex for NGC~4755 and $-0.05\pm0.05$~dex for NGC~6231.
These values are consistent with our previous results (Daszy\'nska et al. \cite{daszynska02}),
obtained with the less accurate method.

The values of the metallicity parameter we obtained, are not correlated with any pulsational parameter
or the projected rotational velocity. However, we have noticed that, on the whole,
multiperiodic variables have higher values of metallicity than the monoperiodic ones. 
This is consistent with pulsation theory.

The abundance of metals, especially that of iron, in $\beta$~Cephei variables is one of the fundamental parameters. 
In the case of main-sequence B stars, metallicities, [m/H], obtained from the UV spectra give mainly an estimation 
of [Fe/H]. 

The value of [Fe/H] in the Kurucz's models is larger
than in the widely used in the pulsation theory OPAL and OP data. One can therefore expect that [Fe/H]
determined using the ATLAS9 models should be lower in comparison with [Fe/H]
obtained on the basis of the OPAL or OP data (see also Fitzpatrick \& Massa \cite{fitzpatrick99}).
To compare observational metallicities of the stars with the values resulting from the theory,
we have to include the correction for [m/H] of about 0.12~dex.
The metallicities of the $\beta$~Cep stars predicted by the theory of pulsations are not in 
contradiction with most values determined by us. 
Pamyatnykh (\cite{pamyatnykh99}) showed that the $\beta$~Cephei instability strip vanishes for $Z \approx 0.01$,
corresponding to [m/H] $\approx -0.30$ dex, and to [m/H] $\approx -0.42$~dex from the Kurucz's fluxes.
There are several stars with metallicity parameter lower than this limit. But we have to remember that our values
give information about photospheric metal abundances.
Some other effects which we did not take into account such as diffusion or element mixing may also be important.
Besides, we used a standard mixture of elements.
Nevertheless, our results provide important information about the metallicity
range for these pulsating stars. In particular, because there is a problem with the estimation
of the [m/H] parameter for early type stars from photometry, making use of low-resolution ultraviolet spectra
seems up to now the best and fastest way to determine the metallicity of B-type pulsating variables.

However, for asteroseismological purposes, there is a need for a detailed
analysis of chemical composition, because oscillation frequencies are very sensitive to the adopted mixture,
and this is the aim of our future work.\\

\noindent {\it Acknowledgements} We thank prof. Dziembowski and prof A. Pamyatnykh for permission of using the evolution
code and non-adiabatic pulsation code.

\begin{table*}
\centering
\caption[]{The list of $\beta$~Cephei stars. The names, HD numbers, the Hipparcos parallax, $\pi$,
numbers of the IUE images used, sources of the ground-based data (last column) are given.}
\begin{scriptsize}
\begin{tabular}{crcllll}
\hline \hline
name           &\multicolumn{1}{c}{HD}    &           &                & IUE numbers                &                            & \\ 
               &\multicolumn{1}{c}{number}&$\pi$ [mas]& SWP camera     & LWP camera                 & LWR camera                 & References\\ \hline \hline
HN Aqr         &         &                &23355, 37224, 37225, 37226, &03661                       &                            &{\bf 6}  \\
               &         &                &37227, 37228, 37229, 37230, &                            &                            &  \\
               &         &                &37231                       &                            &                            &  \\
$\gamma$~Peg   &     886 &$9.79 \pm 0.81$ &52770, 52772, 52775, 52777, &29482, 29483, 29485, 29493, &                            &{\bf 2}, 3, 4, 5, 6  \\
               &         &                &52783                       &29495                       &                            &  \\
$\delta$~Cet   &   16582 &$5.04 \pm 0.83$ &29807, 29808, 29809, 29810, &06341, 09634, 09635, 09636, &                            &{\bf 2}, 3, 4, 5, 6  \\
               &         &                &29811, 29812, 29813, 29814  &09637                       &                            &  \\
$\mu$~Eri      &   29248 &$5.56 \pm 0.88$ &37958, 37959, 37960, 37961, &                            &04142                       &{\bf 2}, 3, 4, 5, 6  \\
               &         &                &37962                       &                            &                            &  \\
$\beta$~CMa    &   44743 &$6.53 \pm 0.66$ &42724, 42725, 42728, 42729, &15384, 04531, 04532         &                            &1, {\bf 2}, 3, 5, 6  \\
               &         &                &42730, 42731, 42732, 42733, &                            &                            &  \\
               &         &                &42734, 42735, 42736         &                            &                            &  \\
$\xi ^1$~CMa   &   46328 &$1.59 \pm 0.70$ &19244                       &                            &15280                       &{\bf 2}, 6  \\
$15$~CMa       &   50707 &$2.02 \pm 0.70$ &03575                       &                            &02426, 02434, 01349         &1, 2, 6  \\
$19$~Mon       &   52918 &$2.92 \pm 0.87$ &21913, 22113, 27665, 27757, &07019                       &                            &{\bf 3}, 4, 5 ,6  \\
               &         &                &29239, 29348, 29385, 30393, &                            &                            &  \\
               &         &                &32115, 32399, 32926, 35104, &                            &                            &  \\
               &         &                &49291, 50412, 52939         &                            &                            &  \\
$27$~CMa       &   56014 &$2.07 \pm 0.59$ &03392, 02866                &                            &02979, 03402, 02543         &{\bf 2}, 6  \\
$\beta$~Cru    &  111123 &$9.25 \pm 0.61$ &05197, 05198, 05199, 05200  &                            &04502                       &{\bf 2}, 6  \\
               &         &                &05201, 05202, 05203, 05204  &                            &                            &  \\
               &         &                &05205                       &                            &                            &  \\
$\alpha$~Vir   &  116658 &$12.44 \pm 0.86$&50032, 18950, 33082, 33091  &12841                       &13650, 15000                &{\bf 2}, 3, 4, 5, 6  \\
$\epsilon$~Cen &  118716 &$8.68 \pm 0.77$ &04347, 05707                &                            &03837                       &{\bf 1}, 2, 6  \\
$\nu$~Cen      &  120307 &$6.87 \pm 0.77$ &36481, 36482                &24753                       &                            &{\bf 1}, 2, 6  \\
$\beta$~Cen    &  122451 &$6.21 \pm 0.56$ &17381                       &                            &13628, 13629                &{\bf 1}, 2, 6  \\
$\tau ^1$~Lup  &  126341 &$3.15 \pm 0.69$ &05206, 05207, 05208         &                            &04503                       &{\bf 2}, 6  \\
$\alpha$~Lup   &  129056 &$5.95 \pm 0.76$ &04564, 04565, 04566, 04567, &23572                       &                            &{\bf 1}, 2, 6  \\
               &         &                &04568, 04569, 04573, 04574  &                            &                            &  \\
BU~Cir         &  129557 &$1.89 \pm 0.69$ &16057, 16058                &                            &12353                       &{\bf 6}  \\
V836~Cen       &  129929 &$1.48 \pm 1.03$ &07687                       &                            &06697                       &{\bf 6}  \\
$\beta$~Lup    &  132058 &$6.23 \pm 0.71$ &16770, 16775, 17457, 17458  &                            &13739                       &{\bf 1}, 6  \\
$\delta$~Lup   &  136298 &$6.39 \pm 0.86$ &16778                       &                            &13034                       &{\bf 1}, 6  \\
$\omega ^1$~Sco&  144470 &$7.70 \pm 0.87$ &09237, 09239, 42228         &20992                       &                            &{\bf 1}, 6  \\
$\sigma$~Sco   &  147165 &$4.44 \pm 0.81$ &45517                       &23843                       &                            &{\bf 2}, 6  \\
$\theta$~Oph   &  157056 &$5.79 \pm 0.69$ &04421, 04422, 04423, 04424, &24025                       &                            &{\bf 1}, 2, 6  \\
               &         &                &04425, 04426, 04427, 04428, &                            &                            &  \\
               &         &                &04429, 04430                &                            &                            &  \\
$\lambda$~Sco  &  158926 &$4.64 \pm 0.90$ &06270, 17385, 17398, 20627  &12737, 12738, 12739, 12740, &                            &{\bf 1}, 2, 6  \\
               &         &                                             &12741, 12742, 12743         &                            &  \\
$\kappa$~Sco   &  160578 &$7.03 \pm 0.73$ &04323, 04348                &                            &03839                       &{\bf 1}, 6  \\
V2052~Oph      &  163472 &$3.93 \pm 0.97$ &50411, 50431, 50642, 50644, &                            &11089                       &{\bf 1}, 6  \\
               &         &                &50651, 50658, 50659, 52122, &                            &                            &  \\
               &         &                &55480, 55488, 55506, 55526, &                            &                            &  \\
               &         &                &55531, 55535, 55540, 55571, &                            &                            &  \\
               &         &                &55601, 55610, 55633, 55648, &                            &                            &  \\
               &         &                &55654, 55668, 55672, 55683, &                            &                            &  \\
               &         &                &55688, 55698, 55702, 55711, &                            &                            &  \\
               &         &                &55714, 55732, 55737, 55746, &                            &                            &  \\
               &         &                &55753, 55757, 55782, 55783, &                            &                            &  \\
               &         &                &50636, 50639, 18313, 14514, &                            &                            &  \\
               &         &                &14515                       &                            &                            &  \\
BW~Vul      &  199140 &$1.84\pm 0.68$  &52642, 52643, 52644, 52648, &                            &04843, 04844, 04845, 04846  &{\bf 6}  \\
&           &                &05590, 05591, 05597, 05598, &                            &                            &  \\
&           &                &52847, 52848, 52849, 52850, &                            &                            &  \\
&           &                &52851, 52852, 52853, 52854, &                            &                            &  \\
&           &                &52855, 52856, 52862, 52863, &                            &                            &  \\
&           &                &52864, 52865, 52866, 52867, &                            &                            &  \\
&           &                &52868, 52869, 52870, 52871, &                            &                            &  \\
&           &                &52877, 52878, 52879, 52880, &                            &                            &  \\
&           &                &52881, 52882, 52883, 52884, &                            &                            &  \\
&           &                &52885                       &                            &                            &  \\
SY~Equ      &  203664 &$2.23\pm 1.08$  &07354, 07355, 37035, 37036, &16368, 16369, 28408, 28409, &06346, 06347, 08764         &{\bf 6}  \\
&           &                &51083, 51084, 51085, 51086, &28410                       &                            &  \\
&           &                &10069                       &                            &                            &  \\
$\beta$~Cep &  205021 &$5.48\pm 0.47$  &46186, 46255, 52514, 42785, &19491                       &                            &{\bf 2}, 3, 4, 5, 6  \\
&           &                &42800, 42825, 46153, 52415, &                            &                            &  \\
&           &                &52460, 52488, 52547, 52573, &                            &                            &  \\
&           &                &52594, 52620, 52653, 53911  &                            &                            &  \\
12~Lac      &  214993 &$2.34\pm 0.62$  &42631                       &21420, 31313                &                            &{\bf 6}  \\
16~Lac      &  216916 &$2.71\pm 0.69$  &05354, 05355, 05357, 05358, &                            &04599                       &{\bf 6}  \\
            &         &                &05359, 05360, 05361         &                            &                            &  \\
\hline \hline
\end{tabular}
\begin{list}{}{}
\item References: 1 -- Alekseeva et al. (1996), 2 -- Breger (1976), 3 -- Glushneva et al. (1998),
  4 --  Glushneva et al. (1992), 5 -- Kharitonov et al. (1988), 6 -- Mermilliod et al. (1997).
\end{list}
\end{scriptsize}
\label{enjdtab1}
\end{table*}

\begin{table*}
\centering
\caption[2]{The list of $\beta$~Cephei stars belonging to open clusters. ID numbers, names and
numbers of the used IUE images are given.}
\begin{tabular}{rcrr}
\hline
ID Number      &\multicolumn{1}{c}{Name}& \multicolumn{2}{c}{IUE numbers}   \\ 
               &\multicolumn{1}{c}{}    & SWP camera  & LWP/LWR camera  \\ \hline
N3293-10       & V401~Car               &20306, 23759 &16221 \\
N3293-11       &                        &21527        &16996 \\
N3293-14       & V405~Car               &21513, 23568 &16995 \\
N3293-27       & V380~Car               &20307, 23758 &16228 \\
N3293-65       & V412~Car               &20324        &16248 \\
N3293-23       & V404~Car               &21529, 23570 &16998 \\
\hline 
N4755-III-1    & CT~Cru                 &33701        &13352 \\
N4755-I-13     &                        &33702        &13355 \\
N4755-I        & CV~Cru                 &33703        &13356 \\
N4755-G        & BS~Cru                 &33709        &13365 \\
N4755-F        & BW~Cru                 &36249        &15505 \\
\hline 
N6231-110      & V947~Sco               &02762        &02464 \\
N6231-150      & V920~Sco               &24120        &04503 \\
N6231-238      & V964~Sco               &16593        &12829 \\
N6231-261      & V946~Sco               &24119        &04502 \\
N6231-282      & V1032~Sco              &16606        &12846 \\
\hline 
\end{tabular}
\label{enjdtab2}
\end{table*}

\begin{table*}
\centering
\caption[]{The best-fit parameters for the $\beta$~Cephei stars obtained from the IUE/NEWSIPS
           low-resolution data.}
\begin{tabular}{rcccrcr}
\hline
HD      &  Name &$\log T_{\rm eff}$      &  $\log g$ & \multicolumn{1}{c}{[m/H]}&   $\theta$~[mas]  & \multicolumn{1}{c}{$E(B-V)$~[mag]} \\ \hline
 --     &HN~Aqr         &$ 4.332 \pm   0.018 $&$ 4.10 $&$   0.21 \pm   0.09 $&$   0.009 \pm   0.000 $&$   0.068 \pm   0.010 $\\
   886  &$\gamma$~Peg   &$ 4.342 \pm   0.011 $&$ 3.82 $&$  -0.04 \pm   0.08 $&$   0.427 \pm   0.011 $&$   0.010 \pm   0.005 $\\
 16582  &$\delta$~Cet   &$ 4.339 \pm   0.014 $&$ 3.73 $&$  -0.24 \pm   0.09 $&$   0.246 \pm   0.009 $&$   0.014 \pm   0.006 $\\
 29248  &$\mu$~Eri      &$ 4.373 \pm   0.016 $&$ 3.78 $&$   0.05 \pm   0.09 $&$   0.257 \pm   0.011 $&$   0.059 \pm   0.010 $\\
 44743  &$\beta$~CMa    &$ 4.392 \pm   0.014 $&$ 3.74 $&$   0.04 \pm   0.10 $&$   0.581 \pm   0.023 $&$   0.023 \pm   0.009 $\\
 46328  &$\xi^1$~Cma    &$ 4.383 \pm   0.021 $&$ 3.89 $&$  -0.33 \pm   0.19 $&$   0.197 \pm   0.011 $&$  -0.003 \pm   0.010 $\\
 50707  &15~CMa         &$ 4.391 \pm   0.018 $&$ 3.81 $&$   0.05 \pm   0.10 $&$   0.160 \pm   0.008 $&$   0.041 \pm   0.010 $\\
 52918  &19~Mon         &$ 4.380 \pm   0.015 $&$ 3.96 $&$   0.15 \pm   0.10 $&$   0.156 \pm   0.006 $&$   0.046 \pm   0.009 $\\
 56014  &27~CMa         &$ 4.397 \pm   0.022 $&$ 4.18 $&$  -0.47 \pm   0.18 $&$   0.173 \pm   0.010 $&$   0.087 \pm   0.012 $\\
111123  &$\beta$~Cru    &$ 4.413 \pm   0.020 $&$ 3.75 $&$  -0.14 \pm   0.13 $&$   0.752 \pm   0.040 $&$  -0.002 \pm   0.011 $\\
116658  &$\alpha$~Vir   &$ 4.392 \pm   0.012 $&$ 3.70 $&$  -0.12 \pm   0.08 $&$   0.921 \pm   0.031 $&$   0.020 \pm   0.005 $\\
118716  &$\epsilon$~Cen &$ 4.387 \pm   0.023 $&$ 3.72 $&$  -0.14 \pm   0.10 $&$   0.504 \pm   0.029 $&$   0.019 \pm   0.012 $\\
120307  &$\nu$~Cen      &$ 4.360 \pm   0.017 $&$ 3.85 $&$  -0.26 \pm   0.10 $&$   0.322 \pm   0.013 $&$   0.013 \pm   0.007 $\\
122451  &$\beta$~Cen    &$ 4.408 \pm   0.020 $&$ 3.89 $&$  -0.03 \pm   0.15 $&$   1.074 \pm   0.059 $&$   0.028 \pm   0.009 $\\
126341  &$\tau^1$~Lup   &$ 4.356 \pm   0.023 $&$ 3.74 $&$  -0.39 \pm   0.16 $&$   0.217 \pm   0.010 $&$   0.101 \pm   0.026 $\\
129056  &$\alpha$~Lup   &$ 4.357 \pm   0.015 $&$ 3.57 $&$   0.04 \pm   0.10 $&$   0.558 \pm   0.030 $&$   0.045 \pm   0.007 $\\
129557  &BU~Cir         &$ 4.386 \pm   0.028 $&$ 3.92 $&$  -0.27 \pm   0.11 $&$   0.126 \pm   0.028 $&$   0.238 \pm   0.090 $\\
129929  &V836~Cen       &$ 4.380 \pm   0.022 $&$ 3.92 $&$  -0.05 \pm   0.10 $&$   0.040 \pm   0.003 $&$   0.086 \pm   0.010 $\\
132058  &$\beta$~Lup    &$ 4.361 \pm   0.008 $&$ 3.65 $&$  -0.35 \pm   0.11 $&$   0.457 \pm   0.008 $&$   0.039 \pm   0.006 $\\
136298  &$\delta$~Lup   &$ 4.354 \pm   0.010 $&$ 3.69 $&$  -0.25 \pm   0.11 $&$   0.352 \pm   0.007 $&$   0.023 \pm   0.006 $\\
144470  &$\omega^1$~Sco &$ 4.445 \pm   0.010 $&$ 4.11 $&$  -0.45 \pm   0.15 $&$   0.291 \pm   0.006 $&$   0.250 \pm   0.008 $\\
147165  &$\delta$~Sco   &$ 4.410 \pm   0.025 $&$ 3.86 $&$  -0.20 \pm   0.20 $&$   0.597 \pm   0.025 $&$   0.360 \pm   0.032 $\\
157056  &$\theta$~Oph   &$ 4.347 \pm   0.016 $&$ 3.77 $&$  -0.15 \pm   0.12 $&$   0.350 \pm   0.014 $&$   0.017 \pm   0.008 $\\
158926  &$\lambda$~Sco  &$ 4.381 \pm   0.011 $&$ 3.69 $&$  -0.21 \pm   0.08 $&$   0.731 \pm   0.022 $&$   0.048 \pm   0.006 $\\
160578  &$\kappa$~Sco   &$ 4.365 \pm   0.010 $&$ 3.62 $&$  -0.19 \pm   0.10 $&$   0.501 \pm   0.011 $&$   0.025 \pm   0.007 $\\
163472  &V2052~Oph      &$ 4.368 \pm   0.012 $&$ 3.89 $&$  -0.25 \pm   0.16 $&$   0.177 \pm   0.003 $&$   0.393 \pm   0.011 $\\
199140  &BW~Vul         &$ 4.362 \pm   0.017 $&$ 3.71 $&$   0.07 \pm   0.12 $&$   0.093 \pm   0.020 $&$   0.145 \pm   0.079 $\\
203664  &SY~Equ         &$ 4.379 \pm   0.036 $&$ 3.90 $&$  -0.01 \pm   0.21 $&$   0.031 \pm   0.004 $&$   0.060 \pm   0.014 $\\
205021  &$\beta$~Cep    &$ 4.383 \pm   0.015 $&$ 3.69 $&$  -0.07 \pm   0.10 $&$   0.320 \pm   0.013 $&$   0.002 \pm   0.007 $\\
214993  &12~Lac         &$ 4.373 \pm   0.020 $&$ 3.65 $&$  -0.20 \pm   0.10 $&$   0.175 \pm   0.041 $&$   0.181 \pm   0.089 $\\
216916  &16~Lac         &$ 4.359 \pm   0.033 $&$ 3.90 $&$  -0.13 \pm   0.13 $&$   0.137 \pm   0.013 $&$   0.118 \pm   0.018 $\\ \hline
N3293-10  &V401~Car &$   4.370 \pm   0.050 $&$   3.91 $&$  -0.14 \pm   0.12 $&$   0.027 \pm   0.003 $&$   0.276 \pm   0.027 $\\
N3293-11  & --      &$   4.347 \pm   0.022 $&$   3.84 $&$  -0.10 \pm   0.12 $&$   0.023 \pm   0.002 $&$   0.211 \pm   0.012 $\\
N3293-14  &V405~Car &$   4.352 \pm   0.023 $&$   4.18 $&$   0.15 \pm   0.11 $&$   0.027 \pm   0.002 $&$   0.171 \pm   0.010 $\\
N3293-23  &V404~Car &$   4.340 \pm   0.026 $&$   3.94 $&$   0.27 \pm   0.06 $&$   0.075 \pm   0.005 $&$   0.324 \pm   0.016 $\\
N3293-27  &V380~Car &$   4.393 \pm   0.043 $&$   3.91 $&$   0.06 \pm   0.10 $&$   0.039 \pm   0.005 $&$   0.318 \pm   0.019 $\\
N3293-65  &V412~Car &$   4.354 \pm   0.034 $&$   3.97 $&$   0.03 \pm   0.11 $&$   0.022 \pm   0.002 $&$   0.228 \pm   0.016 $\\
\hline
N4755-F    &BW~Cru &$   4.397 \pm   0.010 $&$   3.67 $&$  -0.31 \pm   0.15 $&$   0.040 \pm   0.002 $&$   0.409 \pm   0.009 $\\
N4755-G    &BS~Cru &$   4.439 \pm   0.020 $&$   3.87 $&$  -0.56 \pm   0.20 $&$   0.027 \pm   0.002 $&$   0.432 \pm   0.015 $\\
N4755-I    &CV~Cru &$   4.393 \pm   0.016 $&$   3.68 $&$  -0.29 \pm   0.10 $&$   0.033 \pm   0.002 $&$   0.569 \pm   0.016 $\\
N4755-I-13 & --    &$   4.423 \pm   0.013 $&$   4.03 $&$  -0.49 \pm   0.12 $&$   0.024 \pm   0.001 $&$   0.463 \pm   0.012 $\\
N4755-III-1&CT~Cru &$   4.411 \pm   0.013 $&$   3.74 $&$  -0.50 \pm   0.18 $&$   0.031 \pm   0.002 $&$   0.437 \pm   0.015 $\\
\hline
N6231-110  &V947~Sco  &$   4.400 \pm   0.059 $&$   4.05 $&$  -0.03 \pm   0.23 $&$   0.044 \pm   0.007 $&$   0.495 \pm   0.019 $\\
N6231-150  &V920~Sco  &$   4.409 \pm   0.026 $&$   3.90 $&$  -0.21 \pm   0.19 $&$   0.038 \pm   0.005 $&$   0.459 \pm   0.016 $\\
N6231-238  &V964~Sco  &$   4.438 \pm   0.040 $&$   4.17 $&$   0.05 \pm   0.15 $&$   0.037 \pm   0.005 $&$   0.396 \pm   0.021 $\\
N6231-261  &V946~Sco  &$   4.391 \pm   0.025 $&$   3.95 $&$   0.01 \pm   0.13 $&$   0.028 \pm   0.004 $&$   0.435 \pm   0.015 $\\
N6231-282  &V1032~Sco &$   4.419 \pm   0.028 $&$   3.76 $&$  -0.42 \pm   0.16 $&$   0.038 \pm   0.003 $&$   0.372 \pm   0.012 $\\
\hline
\end{tabular}
\label{enjdtab3}
\end{table*}

\begin{table*}
\centering
\caption[]{The best-fit parameters specifying the shape of the UV extinction curve and parameters of the standard extinction
curve (Fitzpatrick (\cite{fitzpatrick}).}
\begin{tabular}{rccccc}
\hline
HD & \multicolumn{1}{c}{$\gamma$~[$\mu$m$^{-1}$]} &\multicolumn{1}{c}{$x_0$~[$\mu$m$^{-1}$]} & \multicolumn{1}{c}{$c_2$}& \multicolumn{1}{c}{$c_3$} & \multicolumn{1}{c}{$c_4$} \\ \hline \hline
          &$ 0.99         $&$ 4.596           $&$ 0.698           $&$ 3.23          $&$ 0.41          $\\ \hline
126341    &$ 0.72\pm 0.13 $&$ 4.542 \pm 0.031 $&$ 0.893 \pm 0.377 $&$ 1.86 \pm 0.72 $&$ 0.03 \pm 0.25 $\\
129557    &$ 0.82\pm 0.05 $&$ 4.580 \pm 0.010 $&$ 0.579 \pm 0.315 $&$ 2.66 \pm 1.30 $&$ 0.11 \pm 0.18 $\\
144470    &$ 0.87\pm 0.06 $&$ 4.582 \pm 0.012 $&$ 0.866 \pm 0.108 $&$ 2.90 \pm 0.33 $&$ 0.15 \pm 0.10 $\\
147165    &$ 0.94\pm 0.04 $&$ 4.572 \pm 0.007 $&$ 0.326 \pm 0.067 $&$ 3.20 \pm 0.25 $&$ 0.19 \pm 0.06 $\\
163472    &$ 0.65\pm 0.03 $&$ 4.539 \pm 0.007 $&$ 0.562 \pm 0.073 $&$ 2.22 \pm 0.17 $&$ 0.18 \pm 0.11 $\\
199140    &$ 0.77\pm 0.08 $&$ 4.611 \pm 0.021 $&$ 0.875 \pm 0.570 $&$ 1.97 \pm 1.29 $&$ 0.20 \pm 0.11 $\\
214993    &$ 0.86\pm 0.08 $&$ 4.606 \pm 0.024 $&$ 0.776 \pm 0.327 $&$ 1.75 \pm 0.79 $&$ 0.10 \pm 0.21 $\\
\hline
\end{tabular}
\label{enjdtab4}
\end{table*}

\begin{table*}
\centering
\caption[]{The mean correlations, $\rho_{\rm mean}$, their standard deviations, $\sigma_\rho$, median values,
$\rho_m$, and ranges of correlation coefficients between the determined parameters.}
\begin{tabular}{llllll}
\hline
$\rho(p_i,p_j)$ & \multicolumn{1}{c}{$\rho_{mean}$}   &\multicolumn{1}{c}{$\rho_m$} & \multicolumn{1}{c}{$\sigma_\rho$} & \multicolumn{2}{c}{Range of $\rho$}  \\ \hline
$\rho(\theta        -T_{\rm eff})$& 0.657 &   0.784 &   0.296&  0.011&  0.972\\
$\rho(\theta        -E(B-V)     )$& 0.443 &   0.435 &   0.271&  0.001&  0.963\\
$\rho(\theta        -{\rm [m/H]})$& 0.226 &   0.185 &   0.170&  0.002&  0.656\\
$\rho(\theta        -S_1        )$& 0.525 &   0.605 &   0.274&  0.001&  0.996\\
$\rho(\theta        -S_2        )$& 0.579 &   0.696 &   0.292&  0.004&  0.996\\ \hline
$\rho(T_{\rm eff}   -\theta     )$& 0.657 &   0.784 &   0.296&  0.011&  0.972\\
$\rho(T_{\rm eff}   -E(B-V)     )$& 0.726 &   0.768 &   0.184&  0.020&  0.959\\
$\rho(T_{\rm eff}   -{\rm [m/H]})$& 0.261 &   0.217 &   0.189&  0.001&  0.749\\
$\rho(T_{\rm eff}   -S_1        )$& 0.577 &   0.668 &   0.248&  0.029&  0.945\\
$\rho(T_{\rm eff}   -S_2        )$& 0.500 &   0.534 &   0.223&  0.039&  0.880\\ \hline
$\rho(E(B-V)        -\theta     )$& 0.443 &   0.435 &   0.271&  0.001&  0.963\\
$\rho(E(B-V)        -T_{\rm eff})$& 0.726 &   0.768 &   0.184&  0.020&  0.959\\
$\rho(E(B-V)        -{\rm [m/H]})$& 0.186 &   0.157 &   0.141&  0.002&  0.794\\
$\rho(E(B-V)        -S_1        )$& 0.333 &   0.305 &   0.207&  0.007&  0.862\\
$\rho(E(B-V)        -S_2        )$& 0.271 &   0.204 &   0.212&  0.005&  0.859\\ \hline
$\rho({\rm [m/H]}    -\theta     )$& 0.226 &   0.185 &   0.170&  0.002&  0.656\\
$\rho({\rm [m/H]}    -T_{\rm eff})$& 0.261 &   0.217 &   0.189&  0.001&  0.749\\
$\rho({\rm [m/H]}    -E(B-V)     )$& 0.186 &   0.157 &   0.141&  0.002&  0.794\\
$\rho({\rm [m/H]}    -S_1        )$& 0.267 &   0.195 &   0.225&  0.003&  0.881\\
$\rho({\rm [m/H]}    -S_2        )$& 0.166 &   0.128 &   0.129&  0.001&  0.549\\ \hline
$\rho(S_1           -\theta     )$& 0.525 &   0.605 &   0.274&  0.001&  0.996\\
$\rho(S_1           -T_{\rm eff})$& 0.577 &   0.668 &   0.248&  0.029&  0.945\\
$\rho(S_1           -E(B-V)     )$& 0.333 &   0.305 &   0.207&  0.007&  0.862\\
$\rho(S_1           -{\rm [m/H]} )$& 0.267 &   0.195 &   0.225&  0.003&  0.881\\
$\rho(S_1           -S_2        )$& 0.860 &   0.886 &   0.116&  0.500&  1.000\\ \hline
$\rho(S_2           -\theta     )$& 0.579 &   0.696 &   0.292&  0.004&  0.996\\
$\rho(S_2           -T_{\rm eff})$& 0.500 &   0.534 &   0.223&  0.039&  0.880\\
$\rho(S_2           -E(B-V)     )$& 0.271 &   0.204 &   0.212&  0.005&  0.859\\
$\rho(S_2           -{\rm [m/H]} )$& 0.166 &   0.128 &   0.129&  0.001&  0.549\\ 
$\rho(S_2           -S_1        )$& 0.860 &   0.886 &   0.116&  0.500&  1.000\\ \hline
\end{tabular}
\label{enjdtab5}
\end{table*}

\end{document}